\documentclass[onecolumn]{aastex63}
\usepackage{natbib,amsmath,scalerel,placeins}

\usepackage{graphicx}
\usepackage{txfonts}
\usepackage{xspace}
\usepackage{color}
\usepackage{url}
\usepackage{hyperref}
\usepackage{comment}
\usepackage[flushleft]{threeparttable}
\usepackage{lineno}

\newcommand {\bc}{\begin {center}}
\newcommand {\ec}{\end {center}}
\newcommand {\be}{\begin {equation}}
\newcommand {\ee}{\end {equation}}
\newcommand {\beq}{\begin {eqnarray}}
\newcommand {\eeq}{\end {eqnarray}}

\newcommand {\ergs}{{\rm erg\ \rm s^{-1}}}
\def\gro{GRO~J1008-57~}

\begin{document}

\title{Relation of cyclotron resonant energy and luminosity in a strongly magnetized neutron star GRO~J1008-57 observed by Insight-HXMT}
\author{X. Chen}
\affiliation{School of Physics and Technology, Wuhan University, Wuhan 430072, China}
\affiliation{WHU-NAOC Joint Center for Astronomy, Wuhan University, Wuhan 430072, China}
\author{W. Wang}
\affiliation{School of Physics and Technology, Wuhan University, Wuhan 430072, China}
\affiliation{WHU-NAOC Joint Center for Astronomy, Wuhan University, Wuhan 430072, China}
\author{Y. M. Tang}
\affiliation{School of Physics and Technology, Wuhan University, Wuhan 430072, China}
\affiliation{WHU-NAOC Joint Center for Astronomy, Wuhan University, Wuhan 430072, China}
\author{Y. Z. Ding}
\affiliation{School of Physics and Technology, Wuhan University, Wuhan 430072, China}
\affiliation{WHU-NAOC Joint Center for Astronomy, Wuhan University, Wuhan 430072, China}
\author{Y. L. Tuo}
\affiliation{Key Laboratory of Particle Astrophysics, Institute of High Energy Physics, Chinese Academy of Sciences, Beijing 100049, China}
\affiliation{University of Chinese Academy of Sciences, Chinese Academy of Sciences, Beijing 100049, China}
\author{A. A. Mushtukov}
\affiliation{Leiden Observatory, Leiden University, NL-2300RA Leiden, The Netherlands}
\affiliation{Space Research Institute of the Russian Academy of Sciences, Profsoyuznaya Str. 84/32, Moscow 117997, Russia}
\affiliation{Pulkovo Observatory, Russian Academy of Sciences, Saint Petersburg 196140, Russia}
\author{O. Nishimura}
\affiliation{Department of Electronics and Computer Science, Nagano National College of Technology, 716 Tokuma, Nagano 381-8550, Japan}
\author{S. N. Zhang}
\affiliation{Key Laboratory of Particle Astrophysics, Institute of High Energy Physics, Chinese Academy of Sciences, Beijing 100049, China}
\affiliation{University of Chinese Academy of Sciences, Chinese Academy of Sciences, Beijing 100049, China}
\author{M. Y. Ge}
\affiliation{Key Laboratory of Particle Astrophysics, Institute of High Energy Physics, Chinese Academy of Sciences, Beijing 100049, China}
\author{L. M. Song}
\affiliation{Key Laboratory of Particle Astrophysics, Institute of High Energy Physics, Chinese Academy of Sciences, Beijing 100049, China}
\author{F. J. Lu}
\affiliation{Key Laboratory of Particle Astrophysics, Institute of High Energy Physics, Chinese Academy of Sciences, Beijing 100049, China}
\author{S. Zhang}
\affiliation{Key Laboratory of Particle Astrophysics, Institute of High Energy Physics, Chinese Academy of Sciences, Beijing 100049, China}
\author{J. L. Qu}
\affiliation{Key Laboratory of Particle Astrophysics, Institute of High Energy Physics, Chinese Academy of Sciences, Beijing 100049, China}
\correspondingauthor{Wei Wang}
\email{wangwei2017@whu.edu.cn}

\begin{abstract}

Cyclotron line scattering features are detected in a few tens of X-ray pulsars (XRPs) and used as direct indicators of a strong magnetic field at the surface of accreting neutron stars (NSs). In a few cases, cyclotron lines are known to be variable with accretion luminosity of XRPs. It is accepted that the observed variations of cyclotron line scattering features are related to variations of geometry and dynamics of accretion flow above the magnetic poles of a NS. A positive correlation between the line centroid energy and luminosity is typical for sub-critical XRPs, where the accretion results in hot spots at the magnetic poles. The negative correlation was proposed to be a specific feature of bright super-critical XRPs, where radiation pressure supports accretion columns above the stellar surface. Cyclotron line in spectra of Be-transient X-ray pulsar GRO J1008-57 is detected at energies from $\sim 75 -90$ keV, the highest observed energy of cyclotron line feature in XRPs. We report the peculiar relation of cyclotron line centroid energies with luminosity in GRO J1008-57 during the Type II outburst in August 2017 observed by Insight-HXMT. The cyclotron line energy was detected to be negatively correlated with the luminosity at $3.2\times 10^{37}\,\ergs<L<4.2\times 10^{37}\,\ergs$, and positively correlated at $L\gtrsim 5\times 10^{37}\,\ergs$. We speculate that the observed peculiar behavior of a cyclotron line would be due to variations of accretion channel geometry.

\end{abstract}

\keywords{X-rays: binaries -- X-rays: stars -- neutron stars -- Be stars; stars: individual: GRO J1008-57}

\section{Introduction}
\label{sec:Intro}

Classical X-ray pulsars (XRPs) are accreting strongly magnetized neutron stars (NSs) in close binary systems (see \citealt{2015A&ARv..23....2W} for review).
The magnetic field strength in these objects is known to be $\sim 10^{12}\,{\rm G}$. Strong field on a NSs in XRPs affects the geometry of accretion flow.
The accretion disc of stellar wind from a companion star is interrupted at certain distance from a NS (see \citealt{2014EPJWC..6401001L} for review), than the accretion flow is directed by magnetic field towards the small regions located close to the magnetic poles of a NS.
There the kinetic energy of the material turns into heat and is emitted in the form of X-ray radiation.
A strong magnetic field in the vicinity of the NS surface quantizes the energy of electrons (see, e.g., \citealt{2006RPPh...69.2631H}),
and because of that, the Compton scattering and free-free absorption become resonant at the cyclotron energy
\beq
E_{\rm cyc}\approx 11.6\,\left(\frac{B}{10^{12}\,{\rm G}}\right)\,{\rm keV},
\eeq
and its harmonics, where $B$ is the field strength \citep{1986ApJ...309..362D,1991ApJ...374..687H,2016PhRvD..93j5003M}.
The resonances result in cyclotron scattering features in the energy spectra of XRPs.
Because the cyclotron energy is proportional to the magnetic field strength in the line forming regions,
the detection of cyclotron resonant scattering features (CRSFs) in XRPs provides a unique opportunity to measure magnetic field strength of accreting NSs \citep{1978ApJ...219L.105T}.
At present, there are more than 30 XRPs with detected cyclotron scattering features (see \citealt{2019A&A...622A..61S} for review).
The X-ray spectrum of GRO J1008-57 shows the highest fundamental cyclotron absorption line centroid energy among all known XRPs:  the line centroid energy is at $\sim 74 - 90$ keV, which points to the $B$-field strength $\sim 8\times 10^{12}\,{\rm G}$
\citep{2014RAA....14..565W,2014PASJ...66...59Y,2014ApJ...792..108B,2020ApJ...899L..19G}.

Since the discovery of the cyclotron resonant scattering features in XRPs \citep{1978ApJ...219L.105T}, it has been found that the centroid energy of the cyclotron features could vary with the accretion luminosity, pulsation phase and on long (years) time-scales.
In particular, the long-time scale reduction and evolution of the line centroid energy was detected in Her X-1 \citep{2017A&A...606L..13S,2019JHEAp..23...29X}.

The variations of cyclotron line energy with accretion luminosity has been detected in a number of X-ray transients including
V0332+53 \citep{2006MNRAS.371...19T,2010MNRAS.401.1628T},
A0535+26 \citep{2007A&A...465L..21C},
Vela X-1 \citep{2014ApJ...780..133F,2014MNRAS.440.1114W},
GX 304-1 \citep{2012A&A...542L..28K,2017MNRAS.466.2752R},
Cep X-4 \citep{2017A&A...601A.126V},
and 4U 0115+63 \citep{2004ApJ...610..390M,2007AstL...33..368T,2012MNRAS.423.2854L}.
It is remarkable, that the sources of relatively low mass accretion rates show positive correlation between the line centroid energy and luminosity, while the sources of relatively high mass accretion rates show the negative correlation. In one source - V0332+53 \citep{2017MNRAS.466.2143D} - both correlations were reported.
In both transients the positive correlation was detected at low luminosities and the negative one at high luminosity state.
\cite{2016MNRAS.460L..99C} shows that the relation between the line centroid energy and accretion luminosity is not always strict and the same source can show different line centroid energies at the same apparent luminosity (this phenomenon was detected only in one source V0332+53 at present).

Cyclotron lines were shown to be variable with the phase of pulsations in a few XRPs \citep{2015MNRAS.448.2175L,2017MNRAS.466..593L,2016MNRAS.457..258T,2019A&A...621A.134T}.
Recently, it has been found that the cyclotron line can be pulse-phase-transient and appear only at some phase of pulsations in XRPs
\citep{2019ApJ...883L..11M}.

The variability of the cyclotron line centroid energy in the spectra of XRPs is considered to be related to the geometry of accretion flow in close proximity to the surface of a NS.
The geometry of the emitting region is related to the mass accretion rate.
At low mass accretion rates, the radiation pressure is small, and one expects hot spots at the NS surface.
At high mass accretion rates, the radiation pressure is high enough to stop accreting material above the NS surface.
In this case, the accretion column supported by radiation pressure and confined by a strong magnetic field arises above the stellar surface \citep{1976MNRAS.175..395B,1981A&A....93..255W,2015MNRAS.454.2539M}.
The luminosity separating these two accretion regimes is called the ``critical" luminosity $L_{\rm crit}$.
The critical luminosity was shown to be dependent on the magnetic field strength \citep{2015MNRAS.447.1847M}.
The dynamics of the cyclotron line was shown to be dependent on the luminosity state of XRPs (see \citealt{2019A&A...622A..61S} for review).
In particular, the positive correlation between the cyclotron line centroid energy and accretion luminosity is considered to be typical for sub-critical XRPs
\citep{2007A&A...465L..25S, 2012A&A...542L..28K,2014ApJ...780..133F},
while the negative correlation was detected in bright super-critical sources \citep{2004ApJ...610..390M,2006MNRAS.371...19T,2013AstL...39..375B}.
At the same time, there are some sources without any observed correlation between the cyclotron line energy and luminosity \citep{2007A&A...465L..21C}.
Several theoretical models are aiming to explain the variability of a cyclotron line.
The positive correlation was explained by the Doppler effect in accretion channel \citep{2015MNRAS.454.2714M} and alternatively by the onset of collisionless shock above hot spots at low mass accretion rates \citep{1975ApJ...198..671S,2017MNRAS.466.2752R}.
The negative correlation was explained by the variations of accretion column height above the NS surface.
Different models consider different locations of a line forming region at super-critical mass accretion rates, which might be a radiation dominated shock on top of accretion column \citep{2012A&A...544A.123B} or NS surface (see e.g., \citealt{2013ApJ...777..115P,2015MNRAS.448.2175L,2018MNRAS.474.5425M}), which reprocesses a large fraction of beamed radiation from the accretion column.
Alternatively, \cite{2014ApJ...781...30N} argued that some variations of cyclotron line centroid energy could be related to the changes of a beam pattern. By considering
the structure of an accretion column in two dimensions, \cite{2019PASJ...71...42N} suggested that the line-forming region is a region around an accretion
mound in which the bulk velocity in the line-forming region can be considerably different from that in the continuum-forming region which is assumed to be inside an accretion mound, so that the variation of $E_{\rm cyc}$ results from the motion of the accretion mound in the different luminosity ranges.

The Chinese X-ray telescope Insight-HXMT observed the Be/X-ray pulsar GRO J1008-57 during the Type II outburst in August 2017. So that we can study the spectral variations of the X-ray pulsar with different luminosities. In \S 2, The observations of Insight-HXMT on the source GRO J1008-57 was presented in \S 2, and the data and spectral analysis were shown in \S 3. Based on the spectral analysis, the relation between the cyclotron line energy and X-ray luminosity in GRO J1008-57 was studied in \S 4. In \S 5, we presented possible scenarios like the variation of the accretion flow geometry to explain the relationship. Finally, a brief summary was given in \S 6.

\section{Observations of Insight-HXMT}
\label{sec:Observations}

Insight-HXMT consists of three main instruments: High Energy X-ray Telescope (HE, \citealt{2020SCPMA..6349503L}) covering the energy range of 20 - 250 keV with the detection area of 5000 cm$^2$, Medium Energy X-ray Telescope (ME, \citealt{2020SCPMA..6349504C}) covering the energies of 5 - 30 keV with a total detection area of 952 cm$^2$, and Low Energy X-ray Telescope (LE, \citealt{2020SCPMA..6349505C}) covering 1 - 15 keV with the detection area of 384 cm$^2$.
There are three types of Field of View (FoV): 1$^\circ \times 6^\circ$ (FWHM, the small FoV), $6^\circ \times 6^\circ$ (the large FoV), and the blind FoV used to estimate the particle induced instrumental background. Since its launch, Insight-HXMT went through a series of performance verification tests by observing blank sky, standard sources and sources of interest. These tests showed that the entire satellite works smoothly and healthily, and have allowed for the calibration and estimation of the instruments background \citep{2020JHEAp..27...64L}. The Insight-HXMT Data Analysis Software Package (HXMTDAS) V2.02 was used in this work.

GRO J1008-57 was observed during the outburst in August 2017 with the observation time of 292 ks (see Table 1). For the science analysis, we used the following data selection criteria: (1) pointing offset angle $< 0.1^\circ$; (2) pointing direction above Earth $> 10^\circ$; (3) geomagnetic cut-off rigidity value $> 8$; (4) time since SAA passage $> 300$ s and time to next SAA passage $> 300$ s; (5) for LE observations, pointing direction above bright Earth $> 30^\circ$. Then we performed the standard data analysis following the official user guide (see \citealt{2020ApJ...899L..19G}). The observational information of the Insight-HXMT for the outburst is shown in Table 1.

\begin{figure}
	\centering
	\includegraphics[width=9.2cm]{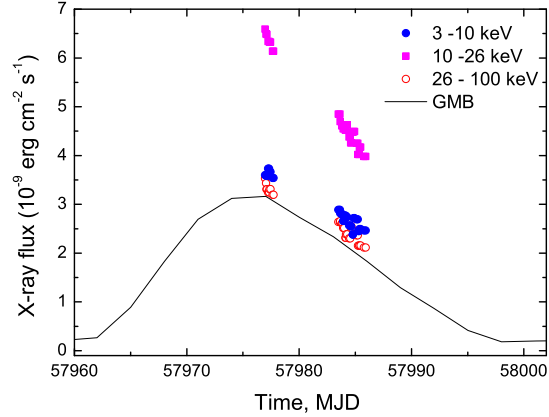}
	\caption{The X-ray light curves of GRO J1008-57 in August 2017. The solid curves obtained from the monitoring data of Fermi/GBM, showed the X-ray outburst lasting about one month. Insight-HXMT has carried out the pointing observations for the outburst, the light curves were presented in three energy bands: 3 - 10 keV, 10 - 26 keV, 26 - 100 keV.
}
\label{fig:light_curves}	
\end{figure}

In Fig.\,\ref{fig:light_curves}, the X-ray light curves of GRO J1008-57 monitored by Insight-HXMT and Fermi/GBM are presented.
The GBM light curves show the type II outburst lasting about one month, the pointing observations performed by Insight-HXMT cover the decay phase of the outburst from the peak luminosity range to about the half of the peak. Thus, the good spectral analysis ability of the Insight-HXMT data covering 1 - 200 keV can for the first time help us to study the spectral variations of GRO J1008-57, including the 90 keV absorption line feature with the decay of the X-ray luminosity, especially probe the variation patterns of the cyclotron line centroid energy with X-ray luminosity.
In the spectral analysis, we have used the energy bands 3 - 10 keV (LE), 10 -26 keV (ME), and 26 - 110 keV (HE), with a systematic error of $1.5\%$ for LE/ME/HE data.

\begin{table}
	\caption{The Insight-HXMT data information used in the analysis taken from the HXMT website: http://www.hxmt.org .}
	\centering
\scriptsize
	\begin{tabular}{@{}cccccc@{}}
		\hline
		Obs. ID & Obs. date Start(UTC) & Obs. date End(UTC) &Duration (s) &  MJD &  \\ \hline
		P0114520001 & 2017-08-11 21:58:37 & 2017-08-12 20:25:06 & 80789 & 57976-57977 &  \\
		P0114520003 & 2017-08-18 11:30:22 & 2017-08-20 22:32:32 & 212530 & 57983-57985 &  \\
\hline
	\end{tabular}
\end{table}

\begin{figure*}
	\centering
	\includegraphics[width=18cm]{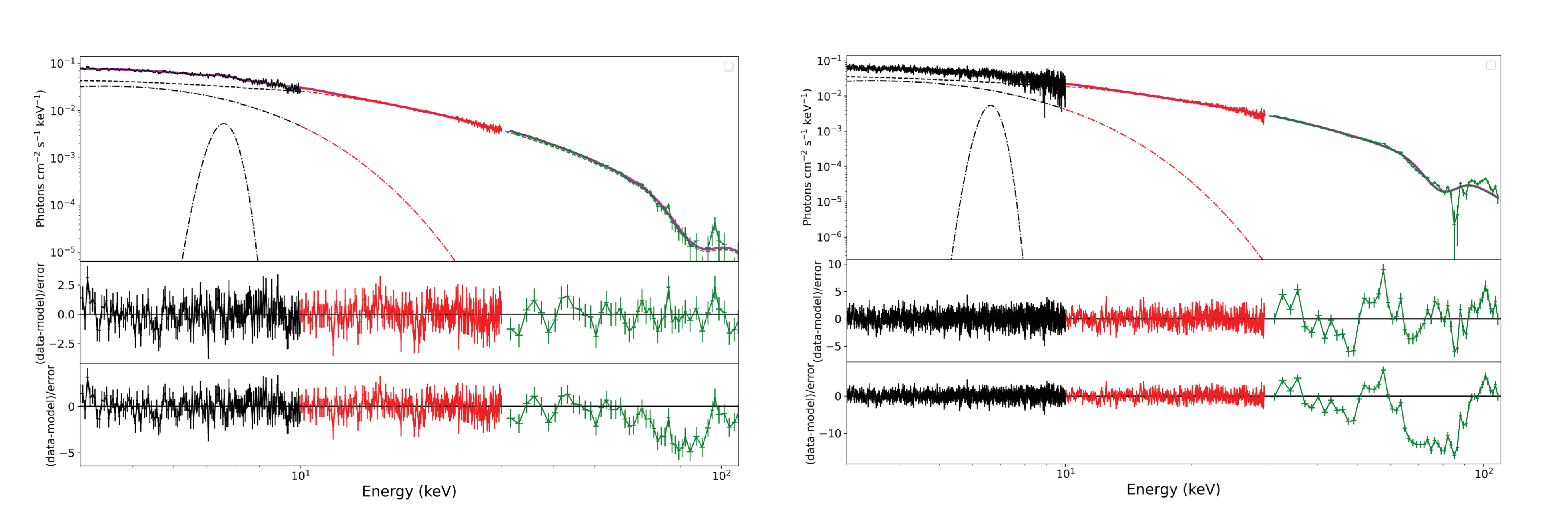}
	\caption{The spectrum examples of GRO J1008-57 from 3 - 110 keV obtained by Insight-HXMT from the peak to decay phase of the outburst. The left panel showed the spectrum around the peak of the outbursts noted with Obs ID 101 in Table 2; the right panel was the spectrum during the decay phase noted with Obs ID 34 in Table 2. For both spectra we have showed the residuals based on the spectral fitting without the cyclotron line component (central panel) and with the cyclotron line (bottom panel), respectively. The spectra were fitted by the Xspec model of power*highecutoff*gabs+bbody+gaussian. }
\label{fig:spectra}
\end{figure*}

\section{Spectral analysis}
\label{sec:Analysis}

Here we have used the Xspec package version 12.10.1 in the spectral analysis work of GRO J1008-57. The spectra of an accreting XRP can be generally described by a power-low model plus the high energy cutoff. Thus, for the X-ray continuum spectral fitting from 3 -- 110 keV, we have used a power-law model plus the high energy cutoff (power*highecutoff) for the first fit. There existed the count excesses below $\sim 5$ keV, we added the thermal component (bbody) to fit the soft X-ray band. After the fittings, the Fe K$\alpha$ emission line around 6.4 keV and the absorption feature around 70 -90 keV appeared in the residuals. We could add the gaussian emission line to fit the Fe K$\alpha$ line ($\sim 6.4$ keV). For the absorption feature above 70 keV, we attributed it to the cyclotron resonant scattering feature, then the model gabs was applied to fit the absorption line features in all observed spectra.

We adopted Monte Carlo Markov Chain (MCMC) method to estimate the errors of the spectral parameters. During the primary run, each chain was generated using 200000 steps (20 walkers) with the first 10000 steps being burnt. During the peak of the outburst, the line width of the cyclotron absorption line was constrained to be around $\sim 14-16$ keV which is still consistent with the early HXMT-Insight result by \cite{2020ApJ...899L..19G}. In the decay phase, the flux became lower, so that the line width can be constrained only for one or two pointing observations, and for the other observations, the line width could not be well constrained. Then to improve the statistics of the spectral fitting in low luminosity range, we combined two neighboring observation IDs into one group for the decay phase, and carried out the spectral fitting again. The cyclotron lines were also detected in the combined spectral analysis, and the line width could be constrained at $\sim 8-11$ keV. All the spectral parameters with the cyclotron line detection for our observations are collected and presented in the Table 2.

In Fig.\,\ref{fig:spectra}, we have shown two spectra as the examples of the spectral fitting results. From the left panel to the right one of Fig.\,\ref{fig:spectra}, the spectra of GRO J1008-57 at the peak and the decay phases of the outburst are presented separately. The residuals are shown for both the cases without the cyclotron line and with the absorption line component.  We also check the possible effect of the different continuum models on the cyclotron line fitting. The NPEX model which is a sum of positive and negative cutoff power-law spectral models \citep{1995PhDT.......215M} could fit the continuum, the cyclotron absorption features also appeared in the residuals (also see \citealt{2020ApJ...899L..19G}), and the determined line energy was consistent with that by the high energy cutoff model. The other continuum models, like BMC, CompTT models cannot fit the continuum well, thus we did not consider them in this work.

\section{Cyclotron line energy versus X-ray luminosity in GRO J1008-57}
\label{sec:Correlation}
\begin{table*}
	\caption{The spectral parameters for the spectra with the cyclotron line detections in GRO J1008-57. During the decay phase, the combined spectra (two pointing observations combined into one Obs ID, i.e., 303 plus 304 as 34, 311 plus 312 as 1122) are fitted. Flux is given in range of 3-100 keV, the corresponding luminosity is also presented with the distance of 5.8 kpc.}
	\centering
\scriptsize
	\begin{tabular}{lcccccccccccl}
		\hline
		Obs ID & kT (keV) & $\Gamma$ & $E_{\rm cut}$ (keV) & $E_{\rm fold}$(keV)  & $E_{\rm cyc}$(keV) & $\sigma$ (keV) & Strength & $E_{Fe}$(keV)& $\sigma_{Fe}$(keV) & Flux (erg cm$^{-2}$ s$^{-1}$) & $L_{\rm x}$(erg s$^{-1}$) & reduced $\chi^2$  \\
\hline
		101 & $1.9\pm 0.1$ & $0.4\pm 0.1$ & $10.8\pm 0.3$ & $14.6\pm 0.4$ & $84.8\pm 1.9$ & $13.9\pm 2.9$ &  $49.6\pm 6.1$ & $6.51\pm0.04$ & $0.25\pm0.05$ & $(1.38\pm 0.02)\times 10^{-8}$& $(5.52\pm 0.04)\times 10^{37}$& 0.905\\
		102 & $2.0\pm 0.1$ & $0.6\pm 0.1$ & $11.1\pm 0.4$ & $16.9\pm 0.4$ & $86.1\pm 2.4$ & $15.1\pm 3.3$ &$47.5\pm 10.5$ &$6.52\pm0.04$ & $0.21\pm0.05$& $(1.38\pm 0.02)\times 10^{-8}$& $(5.52\pm 0.04)\times 10^{37}$& 0.996 \\
        103& $1.9\pm 0.1$ & $0.5\pm 0.1$ & $13.1\pm 0.4$ & $15.6\pm 0.4$ & $91.8\pm 4.9$ & $15.8\pm 3.3$ & $66.1\pm 11.9$ &$6.48\pm0.04$ & $0.23\pm0.05$ &$(1.49\pm 0.02)\times 10^{-8}$ & $(5.96\pm 0.04)\times 10^{37}$ & 0.933 \\
        104 & $1.9\pm 0.1$ & $0.5\pm 0.1$ & $10.9\pm 0.4$ &$16.6\pm 0.4$ & $93.5\pm 5.8$ & $15.5\pm 3.9$ & $85.6\pm 21.5$ &$6.50\pm0.04$ & $0.21\pm0.04$& $(1.44\pm 0.02)\times 10^{-8}$ & $(5.76\pm 0.04)\times 10^{37}$ & 1.012 \\
        105 & $2.1\pm 0.1$ & $0.7\pm 0.1$ & $17.1\pm 0.6$ &$19.2\pm 0.8$ & $86.6\pm 2.5$ & $15.3\pm 3.7$ &$73.3\pm 10.1$ &$6.55\pm0.05$ & $0.23\pm0.06$& $(1.42\pm 0.02)\times 10^{-8}$& $(5.68\pm 0.04)\times 10^{37}$ & 0.947 \\
        106 & $2.0\pm 0.1$ & $0.5\pm 0.1$ & $13.1\pm 0.6$ &$16.9\pm 1.1$ & $82.5\pm 3.4$ & $14.3\pm 3.9$ &$68.5\pm 20.2$ &$6.59\pm0.05$ & $0.13\pm0.06$& $(1.33\pm 0.02)\times 10^{-8}$& $(5.31\pm 0.04)\times 10^{37}$ & 0.959  \\
        12 & $1.6\pm 0.1$ & $0.6\pm 0.1$ & $11.9\pm 0.5$ & $15.8\pm 0.5$ & $76.1\pm 2.6$ &  $8.1\pm 3.1$ &$16.6\pm 7.2$ &$6.42\pm0.07$ & $0.16\pm0.08$& $(1.03\pm 0.02)\times 10^{-9}$ &$(4.15\pm 0.04)\times 10^{37}$ & 1.185 \\
        34 & $1.7\pm 0.1$ & $0.7\pm 0.1$ & $11.2\pm 0.5$ & $17.1\pm 0.6$ & $78.9\pm 2.9$ &  $8.9\pm 2.9$ &$26.1\pm 8.2$ &$6.62\pm0.07$ & $0.28\pm0.09$& $(1.00\pm 0.02)\times 10^{-9}$ &$(4.01\pm 0.04)\times 10^{37}$ & 1.091 \\
       56 & $1.6\pm 0.1$ & $0.7\pm 0.1$ & $11.0\pm 0.4$ & $16.3\pm 0.7$ & $80.7\pm 3.1$ & $9.5\pm 2.9$ &$43.5\pm 14.4$ &$6.66\pm0.07$ & $0.33\pm0.09$& $(9.8\pm 0.2)\times 10^{-9}$ &$(3.89\pm 0.04)\times 10^{37}$ & 1.203  \\
      78 & $1.8\pm 0.1$ & $0.5\pm 0.1$ & $13.7\pm 0.9$ & $15.4\pm 0.7$ & $85.9\pm 3.8$ & $10.2\pm 3.1$ & $51.4\pm 12.8$ &$6.46\pm0.07$ & $0.15\pm0.08$& $9.3\pm 0.2)\times 10^{-9}$ &$(3.77\pm 0.04)\times 10^{37}$ & 1.118  \\
     910 & $1.5\pm 0.1$ & $0.5\pm 0.1$ & $7.9\pm 1.4$ & $14.9\pm 1.3$ & $84.2\pm 4.2$ & $10.1\pm 3.5$  &$21.2\pm 10.9$ &$6.51\pm0.07$ & $0.14\pm0.07$& $9.4\pm 0.2)\times 10^{-9}$ &$(3.81\pm 0.04)\times 10^{37}$ & 1.147 \\
     1112 & $1.6\pm 0.1$ & $0.6\pm 0.1$ & $10.9\pm 0.9$ & $15.6\pm 1.4$ & $83.6\pm 4.6$ &$9.7\pm 3.5$  &$50.9\pm 12.3$ &$6.47\pm0.09$ & $0.18\pm0.07$& $(9.3\pm 0.2)\times 10^{-9}$  &$(3.77\pm 0.04)\times 10^{37}$ & 1.164 \\
     1314 & $1.7\pm 0.1$ & $0.8\pm 0.1$ & $10.9\pm 0.4$ & $15.6\pm 0.4$ & $88.2\pm 3.9$ & $10.1\pm 3.2$ & $57.5\pm 14.6$ &$6.60\pm0.11$ & $0.33\pm0.11$ & $(8.4\pm 0.2)\times 10^{-9}$ &$(3.39\pm 0.04)\times 10^{37}$ &1.232 \\	
     1516 & $1.7\pm 0.1$ & $0.4\pm 0.1$ & $11.0\pm 0.4$ & $14.4\pm 0.4$ & $86.7\pm 3.8$ & $10.0\pm 3.2$ & $20.4\pm 5.6$ &$6.53\pm0.10$ & $0.19\pm0.08$ & $(8.7\pm 0.2)\times 10^{-9}$ &$(3.53\pm 0.04)\times 10^{37}$ &1.203 \\	
     1718 & $1.6\pm 0.1$ & $0.7\pm 0.1$ & $11.4\pm 0.4$ & $16.4\pm 0.5$ & $80.2\pm 3.5$ &  $9.1\pm 3.3$& $21.0\pm 5.9$ &$6.43\pm0.11$ & $0.13\pm0.06$ & $(8.9\pm 0.2)\times 10^{-9}$ &$(3.60\pm 0.04)\times 10^{37}$ & 1.215 \\	

\hline
	\end{tabular}
\end{table*}

\begin{figure*}
	\centering
	\includegraphics[width=10.5cm]{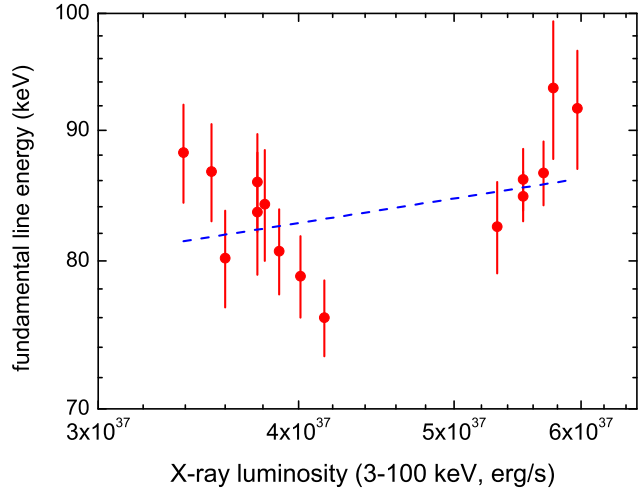}
	\caption{The fundamental line centroid energy versus X-ray luminosity (3 - 100 keV) of GRO J1008-57 in the outburst. The dashed line shows the best fitting power-law function when we assume the positive relation between $E_{\rm cyc}$ and $L_x$. The data points are plotted in the logarithmic scale for both X and Y axes.}	
\label{fig:LCE}
\end{figure*}

The possible variation of the cyclotron line centroid energy in GRO J1008-57 was also reported by Suzaku observations during the 2012 outburst \citep{2014PASJ...66...59Y}, the line energy decreased toward higher X-ray luminosity, however, only three data points were very limited. Based on the spectral analysis results in \S 3, we can collect the fitted line centroid energy and obtained X-ray flux from 3- 100 keV, then study the variation of the cyclotron line energy with X-ray luminosity. In Fig.\,\ref{fig:LCE}, we presented the relation of the fundamental line centroid energy ($E_{\rm cyc}$) versus the X-ray luminosity ($L_x$) of GRO J1008-57, where the X-ray luminosity was calculated based on the estimated distance of 5.8 kpc \citep{2012A&A...539A.114R} and the determined X-ray flux from 3 - 100 keV during the outburst (also see Table 2). The relationships between $E_{\rm cyc}$ and X-ray luminosity are complicated and different from the relations found in other accreting XRPs. In Fig.\,\ref{fig:LCE}, the line centroid energy increases from 80 keV to 90 keV in X-ray luminosity range above $\sim 5\times 10^{37}$ erg/s; but in the lower X-ray luminosity range of $(3.2-4.2)\times 10^{37}$ erg/s, the fundamental line centroid energy varies from $\sim 90$ to 75 keV.

From the Insight-HXMT observations, the fundamental cyclotron line energy of GRO J1008-57 was determined to be $\sim 75-90$ keV, and the magnetic field value higher than $\sim 7\times 10^{12}$ G, which suggests a critical luminosity of $L_{\rm crit}> 9\times 10^{37}$ erg s$^{-1}$. The observed X-ray luminosity ranges of GRO J1008-57 during the outburst were below the critical luminosity. Generally for known XRPs, the variation patterns of the cyclotron line energy show the positive correlation with X-ray luminosity below $L_{\rm crit}$. Thus we would expect a positive correlation of $E_{\rm cyc}-L_x$ in case of GRO J1008-57 using our observations. Thus, we at first used the single power-law function to fit the data points to check the validity of a positive correlation in this source.

For the cyclotron line detection results by the Insight-HXMT observations, we can first fit all the available data points with a function of the single power law form (see  Fig.\,\ref{fig:LCE}). The best fitted function is given by $E_{\rm cyc}\propto L_x^{0.09\pm 0.06}$ with the Pearson coefficient of 0.407, and reduce $\chi^2$ of 1.893 (13 degrees of freedom, hereafter d.o.f). If we consider it as a positive correlation, then the relation should be very weak.

\begin{figure*}
	\centering
	\includegraphics[width=8cm]{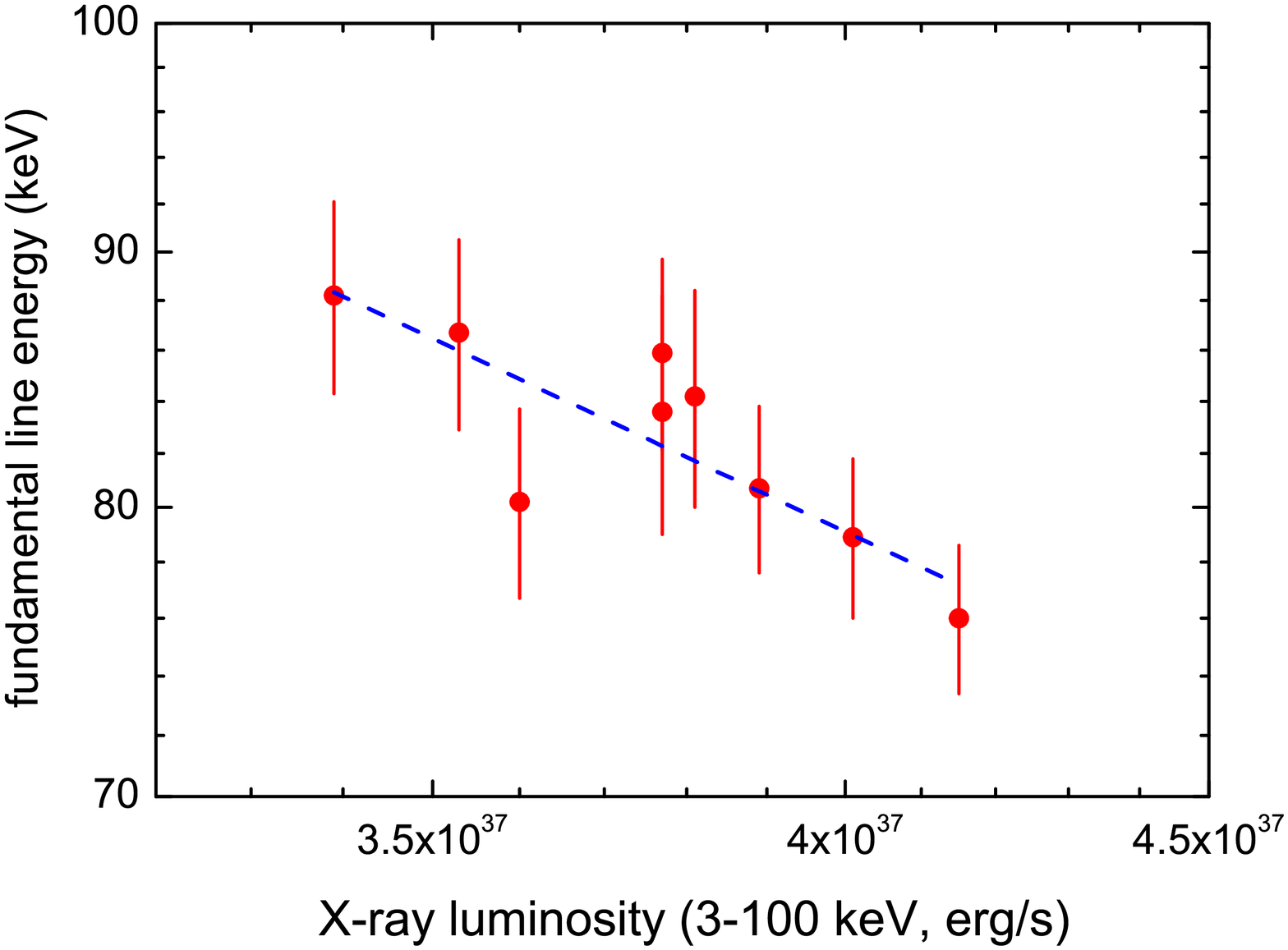}
    \includegraphics[width=8.1cm]{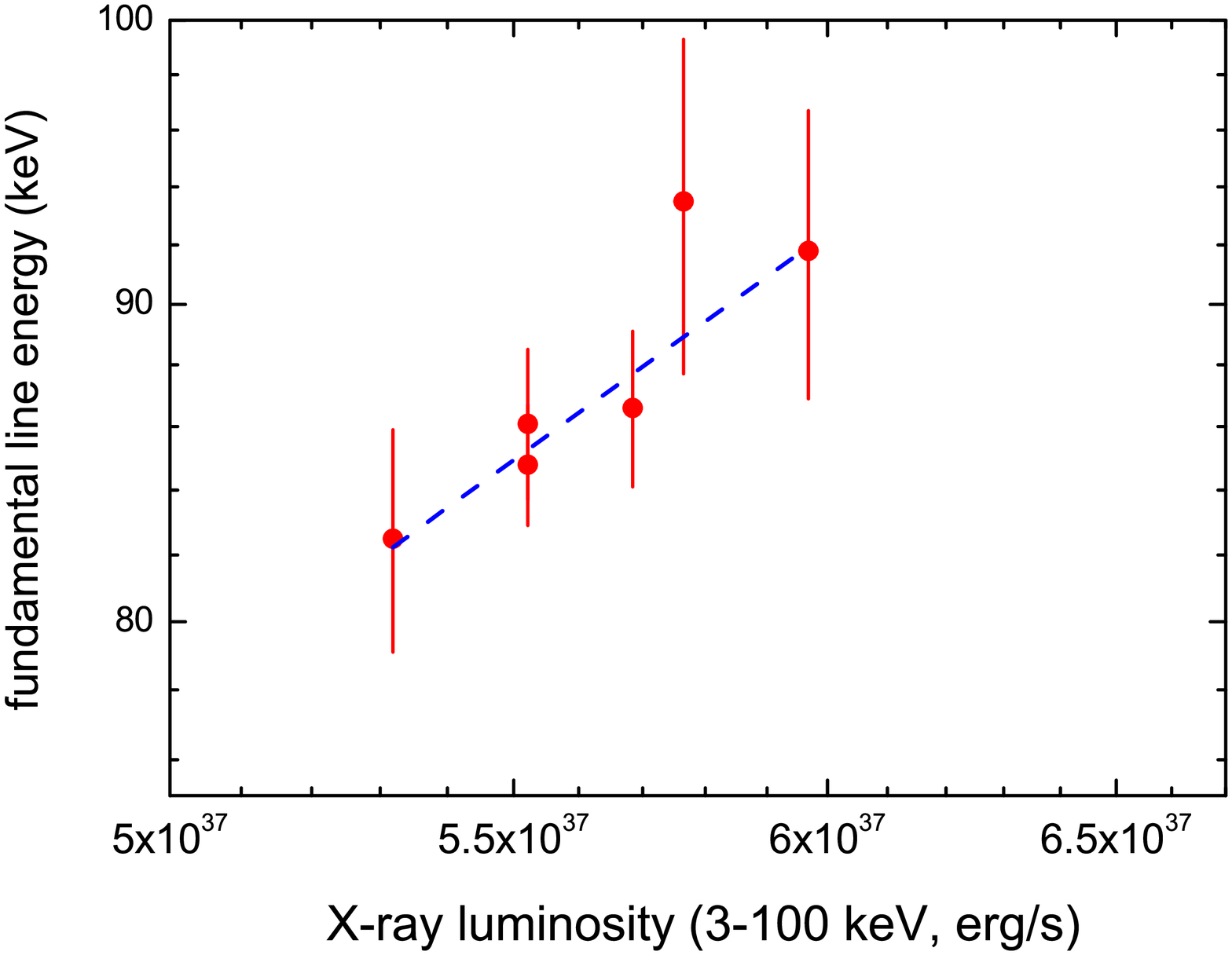}
	\caption{The fundamental line centroid energy versus X-ray luminosity (3 - 100 keV) of GRO J1008-57 for two luminosity ranges. The dashed lines show the best fitting functions with a single power-law form. The left panel presents the negative relation between $E_{\rm cyc} - L_x$ below the luminosity of $\sim 4.2\times 10^{37}$ erg s$^{-1}$; the right panel presents the positive relation between $E_{\rm cyc} - L_x$ above the luminosity of $5\times 10^{37}$ erg s$^{-1}$. The data points are plotted in the logarithmic scale for both X and Y axes. See details in the text.
}	
\label{fig:relation2}
\end{figure*}

For further analyzing the $E_{\rm cyc}-L_x$ relationship in GRO J1008-57, we then fit the data points in two X-ray luminosity ranges.  From the X-ray luminosity range of $(3-5)\times 10^{37}$ erg/s (see the left panel of Fig.\,\ref{fig:relation2}), we fitted the data points with a single power-law function, and obtained a best fitted function of $E_{\rm cyc}\propto L_x^{-0.67\pm 0.15}$ with the Pearson coefficient of $-0.856$, and reduce $\chi^2$ of 1.326 (7 d.o.f). This best fit suggests a negative correlation between $E_{\rm cyc}-L_x$ when $L_x\sim (3.2-4.2)\times 10^{37}$ erg/s. For the higher X-ray luminosity ranges of $>5\times 10^{37}$ erg/s (see the right panel of Fig.\,\ref{fig:relation2}), the best fitted function is given by $E_{\rm cyc}\propto L_x^{0.96\pm 0.25}$ with the Pearson coefficient of 0.885, and reduce $\chi^2$ of 1.126 (4 d.o.f). This positive correlation is much more significant than that with all observed data covering both low and high luminosity ranges.

Thus as a summary, from the X-ray luminosity range of $(3.2-4.2)\times 10^{37}$ erg/s, the fundamental line centroid energy varies from 90 to 75 keV, showing a negative correlation of $E_{\rm cyc}-L_x$, which also confirmed the marginal result found in \cite{2014PASJ...66...59Y}. While in higher luminosity range above $\sim 5\times 10^{37}$ erg/s, the line centroid energy increases from 80 keV to 90 keV, thus showing a positive correlation of $E_{\rm cyc}-L_x$.

\section{Discussion}
\label{sec:Model}

\subsection {Accretion flow geometry and cyclotron line centroid energy}

The details of spectra formation and cyclotron line variability with the luminosity and pulsation phase are determined by the accretion state of XRP.
The critical luminosity $L_{\rm crit}$ is the minimal luminosity which is sufficiently high to stop the accretion flow above the stellar surface due to the radiation pressure \citep{1976MNRAS.175..395B,2015MNRAS.447.1847M}.
The critical luminosity can be roughly estimated as
\beq\label{eq:L_crit}
L_{\rm crit}=4\times 10^{36}\,\frac{m}{R_6}\left( \frac{l}{2\times 10^5\,{\rm cm}} \right)\frac{\sigma_{\rm T}}{\sigma_{\rm eff}}\,\,\ergs,
\eeq
where $m$ is a NS mass in unites of solar masses,
$R_6$ is the NS radius in units of $10^6\,{\rm cm}$,
$l$ is the length of an accretion channel base at the NS surface, $\sigma_{\rm T}$ is the cross section due to Thomson scattering and $\sigma_{\rm eff}$ as the function of the magnetic field is the effective scattering cross section in a strong magnetic field (see e.g., \citealt{1976MNRAS.175..395B}).
Below the critical luminosity, the accretion process results in formation of hot spots at the magnetic poles of a NS.
Above the critical luminosity, the accretion process results in formation of accretion columns confined by a strong magnetic field and supported by radiation pressure \citep{1975PASJ...27..311I,1976MNRAS.175..395B,1981A&A....93..255W}.
The critical luminosity is known to be dependent on the magnetic field strength at the NS surface \citep{2012A&A...544A.123B,2015MNRAS.447.1847M}.
The cyclotron line in \gro is detected at $E>70\,{\rm keV}$, which points to the surface magnetic field strength $B\gtrsim 7\times 10^{12}\,{\rm G}$.
Because of that, the critical luminosity of this source is expected to be $L_{\rm crit} > 9\times 10^{37}\, \ergs$ (see e.g., \citealt{2012A&A...544A.123B,2015MNRAS.447.1847M}).
Thus, the luminosity range of $(3-7)\times 10^{37} \ergs$ in \gro considered in our paper is expected to be below the critical luminosity, and one expects hot spots in the magnetic pole regions.

The radiation pressure at sub-critical luminosity is not enough to stop the accretion flow above the stellar surface, but still it affects the velocity of the accreting material reducing it down to a certain level below the free fall velocity.
The resonant scattering of X-ray photons by the accreting material results in formation of a cyclotron resonant scattering feature redshifted in respect to the local cyclotron line energy. The shift of a cyclotron scattering feature depends on the accretion flow velocity and therefore on the accretion luminosity of XRPs \citep{2015MNRAS.454.2714M}:
the higher the luminosity, the stronger the radiation pressure, the smaller the accretion flow velocity, the smaller the redshift, and the larger the observed line centroid energy.
That is the basic physical picture proposed by \cite{2015MNRAS.454.2714M} to explain observed positive correlation between the cyclotron energy and luminosity in sub-critical XRPs.

\begin{figure}
	\centering
		\includegraphics[width=9cm]{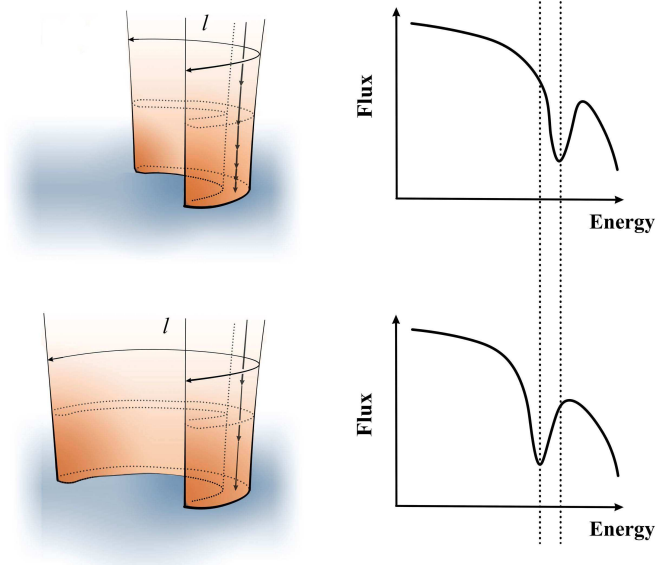}
	\caption{The effectiveness of accretion flow braking is affected by the geometry of accretion flow. The smaller area of accretion channel base at the fixed mass accretion rate results in larger X-ray energy flux in the vicinity of the stellar surface, and, therefore, in stronger radiation pressure. Because the velocity of accreting material near the NS surface influences the cyclotron line centroid energy, the variations of accretion channel geometry affects the displacement of the line in X-ray energy spectrum.}
\label{fig:model_cyc}
\end{figure}

The influence of the radiation pressure on the accretion flow dynamics is determined by the X-ray energy flux near the surface of accreting NS and the braking distance (see Fig.\,\ref{fig:model_cyc}).
The braking distance is close to the geometrical thickness of accretion channel $d$ because at height $h>d$ the X-ray energy flux along magnetic field lines drops sharply due to the delusion effect \citep{1982SvAL....8..330L,1998ApJ...498..790B}.
The X-ray energy flux above the hot spots is determined by the total accretion luminosity, hot spot area $S=ld$ and height $h$ above the surface (see \citealt{2015MNRAS.454.2714M}):
\beq\label{eq:F(h)}
F(h)\approx \frac{L}{2\,ld}\frac{1}{1+(h/d)^2}.
\eeq
The length of accretion channel at the stellar surface can be roughly estimated as
\beq\label{eq:l}
l\sim \xi\,R^{3/2}R_{\rm m}^{-1/2}\,\,{\rm cm},
\eeq
while its geometrical thickness
\beq
d\sim 0.5\,H_{\rm m}R^{3/2}R_{\rm m}^{-3/2}\,\,{\rm cm},
\eeq
where
\beq\label{eq:Rm}
R_{\rm m} \approx 1.3\times 10^{8} B_{12}^{4/7}L_{37}^{-2/7}m^{1/7}R_6^{10/7} \,\,{\rm cm}
\eeq
is the inner radius of accretion disc,
$L_{37}$ is the accretion luminosity in units of $10^{37}\,\ergs$,
$B_{12}$ is magnetic field strength at the poles of a NS in units of $10^{12}\,{\rm G}$,
$H_{\rm m}\ll R_{\rm m}$ is the geometrical thickness of the disc at the inner radius, and
$\xi\in(0;2\pi]$ is the azimutal angle from which magnetized NS in XRPs collects material from the disc at the inner radius (see, e.g., \citealt{2002apa..book.....F}).

The variations of the length $l$ (if any) affect the area of hot spots and result in variations of X-ray energy flux (Eq.\,\ref{eq:F(h)}), the braking effectiveness and the critical luminosity (Eq.\,\ref{eq:L_crit}).
In particular, the decrease of accretion channel length leads to increase of the energy flux and makes the braking of material more effective.

We argue that the observed complicated behaviour of a cyclotron line centroid energy in \gro (see Fig.\,\ref{fig:LCE}) can be a result of variations of accretion channel geometry at luminosity $L\lesssim 4.5\times 10^{37}\,\ergs$.
Decrease of a factor $\xi$ in (Eq.\,\ref{eq:l}) can result in increase of local X-ray energy flux even in the case of total luminosity reduction.
Thus, the positive correlation typical for sub-critical mass accretion rates can turn into the negative one.
The observed switch of positive correlation into the negative one at low mass accretion rates can be explained if the length of the accretion channel base drops by a factor of $\sim 1.7$ with the decrease of the accretion mass.
Because the radius of the open rings at the NS surface is small, the variation of the ring's length will not significantly affect the beam pattern and the observed pulse profile \citep{2020ApJ...899L..19G,2021JHEAp}.

\subsection{Other possible scenarios}
In the framework of the model in \S 5.1, where the cyclotron line forming in accretion channel shifts with the accretion luminosity due to the variations of velocity of accreting material, the line centroid energy is always redshifted in respect to the actual cyclotron line energy.
However, one can expect that at extremely low mass accretion rates the optical thickness of accretion channel is small even for photons experiencing resonant scattering.
In this case, the cyclotron line forms in the atmosphere of a NS and does not experience redshift.
Therefore, the positive correlation between the line centroid energy and luminosity at sub-critical mass accretion rates is expected to turn into the negative one at sufficiently low luminosity.
The switch in the cyclotron line behaviour is expected when the luminosity
\beq
\label{eq:very_low_L}
L \lesssim
\left(\frac{\sigma_{\rm T}}{\sigma_{\rm res}}\right)\,L_{\rm crit} < 10^{-4}L_{\rm crit},
\eeq
where $\sigma_{\rm res}$ is the scattering cross section at the resonance and taken to be $\sigma_{\rm res}>10^4\,\sigma_{\rm T}$ \citep{2016PhRvD..93j5003M}.
In the particular case of \gro the estimation (Eq.\,\ref{eq:very_low_L}) shows that the switch of cyclotron line behaviour due to reduction of optical thickness is expected at $L<10^{34}\,\ergs$.
Therefore, we can exclude this scenario from consideration.

\section{Summary}

We report both the negative and positive correlations between the cyclotron line centroid energies and X-ray luminosity in the Be/X-ray pulsar GRO J1008-57 observed by Insight-HXMT during the outburst in 2017. Because of an extremely strong magnetic field in GRO J1008-57, the XRP is likely in the sub-critical regime of accretion when the radiation pressure at the NS surface is not sufficiently high to stop material above the magnetic poles of a star \citep{1976MNRAS.175..395B,2015MNRAS.447.1847M}. Positive correlation between the line centroid energy and X-ray energy flux detected in the luminosity range $5\times 10^{37}\,\ergs<L<7\times 10^{37}\,\ergs$ is typical for sub-critical XRPs (see, e.g., \citealt{2007A&A...465L..25S,2015MNRAS.454.2714M,2017MNRAS.466.2752R}).
However, the negative correlation observed in GRO J1008-57 in the range $3.2\times 10^{37}\,\ergs < L<4.2\times 10^{37}\,\ergs$ is peculiar for low-luminosity states of accreting strongly magnetized NSs and observed at sub-critical luminosities for the first time. GRO J1008-57 shows the positive correlation turning into the negative one with the decrease of the mass accretion rate. More observational and theoretical work are needed in future.

\section*{Acknowledgements}
We are very grateful to the referee for the useful suggestions to improve the manuscript. This work was supported by the NSFC (U1838103, 11622326, U1838201, U1838202),
the National Program on Key Research and Development Project (Grants No. 2016YFA0400803, 2016YFA0400800), and the Netherlands Organization for Scientific Research Veni Fellowship (AAM). This work made use of data from the Insight-HXMT mission, a project funded by China National Space Administration (CNSA) and the Chinese Academy of Sciences (CAS).

\bibliography{gro1008}

\begin{thebibliography}{}
\expandafter\ifx\csname natexlab\endcsname\relax\def\natexlab#1{#1}\fi
\providecommand{\url}[1]{\href{#1}{#1}}
\providecommand{\dodoi}[1]{doi:~\href{http://doi.org/#1}{\nolinkurl{#1}}}
\providecommand{\doeprint}[1]{\href{http://ascl.net/#1}{\nolinkurl{http://ascl.net/#1}}}
\providecommand{\doarXiv}[1]{\href{https://arxiv.org/abs/#1}{\nolinkurl{https://arxiv.org/abs/#1}}}

\bibitem[{{Basko} \& {Sunyaev}(1976)}]{1976MNRAS.175..395B}
{Basko}, M.~M., \& {Sunyaev}, R.~A. 1976, \mnras, 175, 395,
  \dodoi{10.1093/mnras/175.2.395}

\bibitem[{{Becker}(1998)}]{1998ApJ...498..790B}
{Becker}, P.~A. 1998, \apj, 498, 790, \dodoi{10.1086/305568}

\bibitem[{{Becker} {et~al.}(2012){Becker}, {Klochkov}, {Sch{\"o}nherr},
  {Nishimura}, {Ferrigno}, {Caballero}, {Kretschmar}, {Wolff}, {Wilms}, \&
  {Staubert}}]{2012A&A...544A.123B}
{Becker}, P.~A., {Klochkov}, D., {Sch{\"o}nherr}, G., {et~al.} 2012, \aap, 544,
  A123, \dodoi{10.1051/0004-6361/201219065}

\bibitem[{{Bellm} {et~al.}(2014){Bellm}, {F{\"u}rst}, {Pottschmidt}, {Tomsick},
  {Boggs}, {Chakrabarty}, {Christensen}, {Craig}, {Hailey}, {Harrison},
  {Stern}, {Walton}, {Wilms}, \& {Zhang}}]{2014ApJ...792..108B}
{Bellm}, E.~C., {F{\"u}rst}, F., {Pottschmidt}, K., {et~al.} 2014, \apj, 792,
  108, \dodoi{10.1088/0004-637X/792/2/108}

\bibitem[{{Boldin} {et~al.}(2013){Boldin}, {Tsygankov}, \&
  {Lutovinov}}]{2013AstL...39..375B}
{Boldin}, P.~A., {Tsygankov}, S.~S., \& {Lutovinov}, A.~A. 2013, Astronomy
  Letters, 39, 375, \dodoi{10.1134/S1063773713060029}

\bibitem[{{Caballero} {et~al.}(2007){Caballero}, {Kretschmar}, {Santangelo},
  {Staubert}, {Klochkov}, {Camero}, {Ferrigno}, {Finger}, {Kreykenbohm},
  {McBride}, {Pottschmidt}, {Rothschild}, {Sch{\"o}nherr}, {Segreto}, {Suchy},
  {Wilms}, \& {Wilson}}]{2007A&A...465L..21C}
{Caballero}, I., {Kretschmar}, P., {Santangelo}, A., {et~al.} 2007, \aap, 465,
  L21, \dodoi{10.1051/0004-6361:20067032}

\bibitem[{{Cao} {et~al.}(2020){Cao}, {Jiang}, {Meng}, {Zhang}, {Luo}, {Yang},
  {Zhang}, {Gu}, {Sun}, {Liu}, {Yang}, {Li}, {Tan}, {Liu}, {Du}, {Lu}, {Xu},
  {Guan}, {Zhang}, {Wang}, {Li}, {Zhang}, {Wen}, {Qu}, {Song}, {Li}, {Ge},
  {Zhou}, {Xiong}, {Zhang}, {Zhang}, {Cheng}, {Zhang}, {Li}, {Liang}, {Gao},
  {Yang}, {Liu}, {Liu}, {Yang}, \& {Zhang}}]{2020SCPMA..6349504C}
{Cao}, X., {Jiang}, W., {Meng}, B., {et~al.} 2020, Science China Physics,
  Mechanics, and Astronomy, 63, 249504, \dodoi{10.1007/s11433-019-1506-1}

\bibitem[{{Chen} {et~al.}(2020){Chen}, {Cui}, {Li}, {Wang}, {Xu}, {Lu}, {Wang},
  {Chen}, {Han}, {Hu}, {Zhang}, {Huo}, {Yang}, {Li}, {Lu}, {Zhang}, {Li},
  {Zhang}, {Xiong}, {Zhang}, {Xue}, {Zhao}, {Zhu}, {Zhu}, {Liu}, {Yang}, \&
  {Zhang}}]{2020SCPMA..6349505C}
{Chen}, Y., {Cui}, W., {Li}, W., {et~al.} 2020, Science China Physics,
  Mechanics, and Astronomy, 63, 249505, \dodoi{10.1007/s11433-019-1469-5}

\bibitem[{{Cusumano} {et~al.}(2016){Cusumano}, {La Parola}, {D'A{\`\i}},
  {Segreto}, {Tagliaferri}, {Barthelmy}, \& {Gehrels}}]{2016MNRAS.460L..99C}
{Cusumano}, G., {La Parola}, V., {D'A{\`\i}}, A., {et~al.} 2016, \mnras, 460,
  L99, \dodoi{10.1093/mnrasl/slw084}

\bibitem[{{Daugherty} \& {Harding}(1986)}]{1986ApJ...309..362D}
{Daugherty}, J.~K., \& {Harding}, A.~K. 1986, \apj, 309, 362,
  \dodoi{10.1086/164608}

\bibitem[{{Doroshenko} {et~al.}(2017){Doroshenko}, {Tsygankov}, {Mushtukov},
  {Lutovinov}, {Santangelo}, {Suleimanov}, \& {Poutanen}}]{2017MNRAS.466.2143D}
{Doroshenko}, V., {Tsygankov}, S.~S., {Mushtukov}, A.~A., {et~al.} 2017,
  \mnras, 466, 2143, \dodoi{10.1093/mnras/stw3236}

\bibitem[{{Frank} {et~al.}(2002){Frank}, {King}, \&
  {Raine}}]{2002apa..book.....F}
{Frank}, J., {King}, A., \& {Raine}, D.~J. 2002, {Accretion Power in
  Astrophysics: Third Edition}

\bibitem[{{F{\"u}rst} {et~al.}(2014){F{\"u}rst}, {Pottschmidt}, {Wilms},
  {Tomsick}, {Bachetti}, {Boggs}, {Christensen}, {Craig}, {Grefenstette},
  {Hailey}, {Harrison}, {Madsen}, {Miller}, {Stern}, {Walton}, \&
  {Zhang}}]{2014ApJ...780..133F}
{F{\"u}rst}, F., {Pottschmidt}, K., {Wilms}, J., {et~al.} 2014, \apj, 780, 133,
  \dodoi{10.1088/0004-637X/780/2/133}

\bibitem[{{Ge} {et~al.}(2020){Ge}, {Ji}, {Zhang}, {Santangelo}, {Liu},
  {Doroshenko}, {Staubert}, {Qu}, {Zhang}, {Lu}, {Song}, {Li}, {Tao}, {Xu},
  {Cao}, {Chen}, {Bu}, {Cai}, {Chang}, {Chen}, {Chen}, {Chen}, {Chen}, {Chen},
  {Cui}, {Cui}, {Deng}, {Dong}, {Du}, {Fu}, {Gao}, {Gao}, {Gao}, {Gu}, {Guan},
  {Guo}, {Han}, {Huang}, {Huo}, {Jia}, {Jiang}, {Jiang}, {Jin}, {Jin}, {Kong},
  {Li}, {Li}, {Li}, {Li}, {Li}, {Li}, {Li}, {Li}, {Li}, {Li}, {Liang}, {Liao},
  {Liu}, {Liu}, {Liu}, {Liu}, {Liu}, {Lu}, {Lu}, {Luo}, {Luo}, {Ma}, {Meng},
  {Nang}, {Nie}, {Ou}, {Sai}, {Shang}, {Song}, {Sun}, {Tan}, {Tuo}, {Wang},
  {Wang}, {Wang}, {Wang}, {Wang}, {Wang}, {Wang}, {Wen}, {Wu}, {Wu}, {Wu},
  {Xiao}, {Xiao}, {Xiong}, {Xu}, {Yang}, {Yang}, {Yang}, {Yang}, {Yi}, {Yin},
  {You}, {Zhang}, {Zhang}, {Zhang}, {Zhang}, {Zhang}, {Zhang}, {Zhang},
  {Zhang}, {Zhang}, {Zhang}, {Zhang}, {Zhang}, {Zhang}, {Zhang}, {Zhang},
  {Zhang}, {Zhao}, {Zhao}, {Zheng}, {Zheng}, {Zhou}, {Zhou}, {Zhuang}, {Zhu},
  \& {Zhu}}]{2020ApJ...899L..19G}
{Ge}, M.~Y., {Ji}, L., {Zhang}, S.~N., {et~al.} 2020, \apjl, 899, L19,
  \dodoi{10.3847/2041-8213/abac05}

\bibitem[{{Harding} \& {Daugherty}(1991)}]{1991ApJ...374..687H}
{Harding}, A.~K., \& {Daugherty}, J.~K. 1991, \apj, 374, 687,
  \dodoi{10.1086/170153}

\bibitem[{{Harding} \& {Lai}(2006)}]{2006RPPh...69.2631H}
{Harding}, A.~K., \& {Lai}, D. 2006, Reports on Progress in Physics, 69, 2631,
  \dodoi{10.1088/0034-4885/69/9/R03}

\bibitem[{{Inoue}(1975)}]{1975PASJ...27..311I}
{Inoue}, H. 1975, \pasj, 27, 311

\bibitem[{{Klochkov} {et~al.}(2012){Klochkov}, {Doroshenko}, {Santangelo},
  {Staubert}, {Ferrigno}, {Kretschmar}, {Caballero}, {Wilms}, {Kreykenbohm},
  {Pottschmidt}, {Rothschild}, {Wilson-Hodge}, \&
  {P{\"u}hlhofer}}]{2012A&A...542L..28K}
{Klochkov}, D., {Doroshenko}, V., {Santangelo}, A., {et~al.} 2012, \aap, 542,
  L28, \dodoi{10.1051/0004-6361/201219385}

\bibitem[{{Lai}(2014)}]{2014EPJWC..6401001L}
{Lai}, D. 2014, in European Physical Journal Web of Conferences, Vol.~64,
  European Physical Journal Web of Conferences, 01001,
  \dodoi{10.1051/epjconf/20136401001}

\bibitem[{{Li} {et~al.}(2012){Li}, {Wang}, \& {Zhao}}]{2012MNRAS.423.2854L}
{Li}, J., {Wang}, W., \& {Zhao}, Y. 2012, \mnras, 423, 2854,
  \dodoi{10.1111/j.1365-2966.2012.21096.x}

\bibitem[{{Li} {et~al.}(2020){Li}, {Li}, {Tan}, {Yang}, {Ge}, {Zhang}, {Tuo},
  {Wu}, {Liao}, {Zhang}, {Song}, {Zhang}, {Qu}, {Zhang}, {Lu}, {Xu}, {Liu},
  {Cao}, {Chen}, {Nie}, {Zhao}, \& {Li}}]{2020JHEAp..27...64L}
{Li}, X., {Li}, X., {Tan}, Y., {et~al.} 2020, Journal of High Energy
  Astrophysics, 27, 64, \dodoi{10.1016/j.jheap.2020.02.009}

\bibitem[{{Liu} {et~al.}(2020){Liu}, {Zhang}, {Li}, {Lu}, {Chang}, {Li},
  {Zhang}, {Jin}, {Yu}, {Zhang}, {Fu}, {Chen}, {Ji}, {Xu}, {Deng}, {Shang},
  {Liu}, {Lu}, {Zhang}, {Dong}, {Li}, {Wu}, {Li}, {Wang}, {Wu}, {Zhang},
  {Zhang}, {Xiong}, {Liu}, {Zhang}, {Liu}, {Yang}, \&
  {Zhang}}]{2020SCPMA..6349503L}
{Liu}, C., {Zhang}, Y., {Li}, X., {et~al.} 2020, Science China Physics,
  Mechanics, and Astronomy, 63, 249503, \dodoi{10.1007/s11433-019-1486-x}

\bibitem[{{Lutovinov} {et~al.}(2017){Lutovinov}, {Tsygankov}, {Postnov},
  {Krivonos}, {Molkov}, \& {Tomsick}}]{2017MNRAS.466..593L}
{Lutovinov}, A.~A., {Tsygankov}, S.~S., {Postnov}, K.~A., {et~al.} 2017,
  \mnras, 466, 593, \dodoi{10.1093/mnras/stw3058}

\bibitem[{{Lutovinov} {et~al.}(2015){Lutovinov}, {Tsygankov}, {Suleimanov},
  {Mushtukov}, {Doroshenko}, {Nagirner}, \& {Poutanen}}]{2015MNRAS.448.2175L}
{Lutovinov}, A.~A., {Tsygankov}, S.~S., {Suleimanov}, V.~F., {et~al.} 2015,
  \mnras, 448, 2175, \dodoi{10.1093/mnras/stv125}

\bibitem[{{Lyubarskii} \& {Syunyaev}(1982)}]{1982SvAL....8..330L}
{Lyubarskii}, Y.~E., \& {Syunyaev}, R.~A. 1982, Soviet Astronomy Letters, 8,
  330

\bibitem[{{Mihara}(1995)}]{1995PhDT.......215M}
{Mihara}, T. 1995, PhD thesis, -

\bibitem[{{Mihara} {et~al.}(2004){Mihara}, {Makishima}, \&
  {Nagase}}]{2004ApJ...610..390M}
{Mihara}, T., {Makishima}, K., \& {Nagase}, F. 2004, \apj, 610, 390,
  \dodoi{10.1086/421543}

\bibitem[{{Molkov} {et~al.}(2019){Molkov}, {Lutovinov}, {Tsygankov},
  {Mereminskiy}, \& {Mushtukov}}]{2019ApJ...883L..11M}
{Molkov}, S., {Lutovinov}, A., {Tsygankov}, S., {Mereminskiy}, I., \&
  {Mushtukov}, A. 2019, \apjl, 883, L11, \dodoi{10.3847/2041-8213/ab3e4d}

\bibitem[{{Mushtukov} {et~al.}(2016){Mushtukov}, {Nagirner}, \&
  {Poutanen}}]{2016PhRvD..93j5003M}
{Mushtukov}, A.~A., {Nagirner}, D.~I., \& {Poutanen}, J. 2016, \prd, 93,
  105003, \dodoi{10.1103/PhysRevD.93.105003}

\bibitem[{{Mushtukov} {et~al.}(2015{\natexlab{a}}){Mushtukov}, {Suleimanov},
  {Tsygankov}, \& {Poutanen}}]{2015MNRAS.454.2539M}
{Mushtukov}, A.~A., {Suleimanov}, V.~F., {Tsygankov}, S.~S., \& {Poutanen}, J.
  2015{\natexlab{a}}, \mnras, 454, 2539, \dodoi{10.1093/mnras/stv2087}

\bibitem[{{Mushtukov} {et~al.}(2015{\natexlab{b}}){Mushtukov}, {Suleimanov},
  {Tsygankov}, \& {Poutanen}}]{2015MNRAS.447.1847M}
---. 2015{\natexlab{b}}, \mnras, 447, 1847, \dodoi{10.1093/mnras/stu2484}

\bibitem[{{Mushtukov} {et~al.}(2015{\natexlab{c}}){Mushtukov}, {Tsygankov},
  {Serber}, {Suleimanov}, \& {Poutanen}}]{2015MNRAS.454.2714M}
{Mushtukov}, A.~A., {Tsygankov}, S.~S., {Serber}, A.~V., {Suleimanov}, V.~F.,
  \& {Poutanen}, J. 2015{\natexlab{c}}, \mnras, 454, 2714,
  \dodoi{10.1093/mnras/stv2182}

\bibitem[{{Mushtukov} {et~al.}(2018){Mushtukov}, {Verhagen}, {Tsygankov}, {van
  der Klis}, {Lutovinov}, \& {Larchenkova}}]{2018MNRAS.474.5425M}
{Mushtukov}, A.~A., {Verhagen}, P.~A., {Tsygankov}, S.~S., {et~al.} 2018,
  \mnras, 474, 5425, \dodoi{10.1093/mnras/stx2905}

\bibitem[{{Nishimura}(2014)}]{2014ApJ...781...30N}
{Nishimura}, O. 2014, \apj, 781, 30, \dodoi{10.1088/0004-637X/781/1/30}

\bibitem[{{Nishimura}(2019)}]{2019PASJ...71...42N}
---. 2019, \pasj, 71, 42, \dodoi{10.1093/pasj/psz008}

\bibitem[{{Poutanen} {et~al.}(2013){Poutanen}, {Mushtukov}, {Suleimanov},
  {Tsygankov}, {Nagirner}, {Doroshenko}, \& {Lutovinov}}]{2013ApJ...777..115P}
{Poutanen}, J., {Mushtukov}, A.~A., {Suleimanov}, V.~F., {et~al.} 2013, \apj,
  777, 115, \dodoi{10.1088/0004-637X/777/2/115}

\bibitem[{{Riquelme} {et~al.}(2012){Riquelme}, {Torrejón}, \&
  {Negueruela}}]{2012A&A...539A.114R}
{Riquelme}, M.~S., {Torrejón}, J.~M., \& {Negueruela}, I. 2012, \aap, 539,
  A114, \dodoi{10.1051/0004-6361/201117738}

\bibitem[{{Rothschild} {et~al.}(2017){Rothschild}, {K{\"u}hnel}, {Pottschmidt},
  {Hemphill}, {Postnov}, {Gornostaev}, {Shakura}, {F{\"u}rst}, {Wilms},
  {Staubert}, \& {Klochkov}}]{2017MNRAS.466.2752R}
{Rothschild}, R.~E., {K{\"u}hnel}, M., {Pottschmidt}, K., {et~al.} 2017,
  \mnras, 466, 2752, \dodoi{10.1093/mnras/stw3222}

\bibitem[{{Shapiro} \& {Salpeter}(1975)}]{1975ApJ...198..671S}
{Shapiro}, S.~L., \& {Salpeter}, E.~E. 1975, \apj, 198, 671,
  \dodoi{10.1086/153645}

\bibitem[{{Staubert} {et~al.}(2017){Staubert}, {Klochkov}, {F{\"u}rst},
  {Wilms}, {Rothschild}, \& {Harrison}}]{2017A&A...606L..13S}
{Staubert}, R., {Klochkov}, D., {F{\"u}rst}, F., {et~al.} 2017, \aap, 606, L13,
  \dodoi{10.1051/0004-6361/201731927}

\bibitem[{{Staubert} {et~al.}(2007){Staubert}, {Shakura}, {Postnov}, {Wilms},
  {Rothschild}, {Coburn}, {Rodina}, \& {Klochkov}}]{2007A&A...465L..25S}
{Staubert}, R., {Shakura}, N.~I., {Postnov}, K., {et~al.} 2007, \aap, 465, L25,
  \dodoi{10.1051/0004-6361:20077098}

\bibitem[{{Staubert} {et~al.}(2019){Staubert}, {Tr{\"u}mper}, {Kendziorra},
  {Klochkov}, {Postnov}, {Kretschmar}, {Pottschmidt}, {Haberl}, {Rothschild},
  {Santangelo}, {Wilms}, {Kreykenbohm}, \& {F{\"u}rst}}]{2019A&A...622A..61S}
{Staubert}, R., {Tr{\"u}mper}, J., {Kendziorra}, E., {et~al.} 2019, \aap, 622,
  A61, \dodoi{10.1051/0004-6361/201834479}

\bibitem[{{Truemper} {et~al.}(1978){Truemper}, {Pietsch}, {Reppin}, {Voges},
  {Staubert}, \& {Kendziorra}}]{1978ApJ...219L.105T}
{Truemper}, J., {Pietsch}, W., {Reppin}, C., {et~al.} 1978, \apjl, 219, L105,
  \dodoi{10.1086/182617}

\bibitem[{{Tsygankov} {et~al.}(2019){Tsygankov}, {Doroshenko}, {Mushtukov},
  {Lutovinov}, \& {Poutanen}}]{2019A&A...621A.134T}
{Tsygankov}, S.~S., {Doroshenko}, V., {Mushtukov}, A.~A., {Lutovinov}, A.~A.,
  \& {Poutanen}, J. 2019, \aap, 621, A134, \dodoi{10.1051/0004-6361/201833786}

\bibitem[{{Tsygankov} {et~al.}(2006){Tsygankov}, {Lutovinov}, {Churazov}, \&
  {Sunyaev}}]{2006MNRAS.371...19T}
{Tsygankov}, S.~S., {Lutovinov}, A.~A., {Churazov}, E.~M., \& {Sunyaev}, R.~A.
  2006, \mnras, 371, 19, \dodoi{10.1111/j.1365-2966.2006.10610.x}

\bibitem[{{Tsygankov} {et~al.}(2007){Tsygankov}, {Lutovinov}, {Churazov}, \&
  {Sunyaev}}]{2007AstL...33..368T}
---. 2007, Astronomy Letters, 33, 368, \dodoi{10.1134/S1063773707060023}

\bibitem[{{Tsygankov} {et~al.}(2016){Tsygankov}, {Lutovinov}, {Krivonos},
  {Molkov}, {Jenke}, {Finger}, \& {Poutanen}}]{2016MNRAS.457..258T}
{Tsygankov}, S.~S., {Lutovinov}, A.~A., {Krivonos}, R.~A., {et~al.} 2016,
  \mnras, 457, 258, \dodoi{10.1093/mnras/stv2849}

\bibitem[{{Tsygankov} {et~al.}(2010){Tsygankov}, {Lutovinov}, \&
  {Serber}}]{2010MNRAS.401.1628T}
{Tsygankov}, S.~S., {Lutovinov}, A.~A., \& {Serber}, A.~V. 2010, \mnras, 401,
  1628, \dodoi{10.1111/j.1365-2966.2009.15791.x}

\bibitem[{{Vybornov} {et~al.}(2017){Vybornov}, {Klochkov}, {Gornostaev},
  {Postnov}, {Sokolova-Lapa}, {Staubert}, {Pottschmidt}, \&
  {Santangelo}}]{2017A&A...601A.126V}
{Vybornov}, V., {Klochkov}, D., {Gornostaev}, M., {et~al.} 2017, \aap, 601,
  A126, \dodoi{10.1051/0004-6361/201630275}

\bibitem[{{Walter} {et~al.}(2015){Walter}, {Lutovinov}, {Bozzo}, \&
  {Tsygankov}}]{2015A&ARv..23....2W}
{Walter}, R., {Lutovinov}, A.~A., {Bozzo}, E., \& {Tsygankov}, S.~S. 2015,
  \aapr, 23, 2, \dodoi{10.1007/s00159-015-0082-6}

\bibitem[{{Wang}(2014{\natexlab{a}})}]{2014RAA....14..565W}
{Wang}, W. 2014{\natexlab{a}}, Research in Astronomy and Astrophysics, 14, 565,
  \dodoi{10.1088/1674-4527/14/5/006}

\bibitem[{{Wang}(2014{\natexlab{b}})}]{2014MNRAS.440.1114W}
---. 2014{\natexlab{b}}, \mnras, 440, 1114, \dodoi{10.1093/mnras/stu210}

\bibitem[{{Wang} {et~al.}(2021){Wang}, {Tang}, {Ge}, {Zhang}, {Zhang},
  {Santangelo}, {Ducci}, {Liao}, {Guo}, {Li}, {Zhang}, {Qu}, {Lu}, {Li},
  {Song}, {Xu}, {Bu}, {Cai}, {Cao}, {Chang}, {Chen}, {Chen}, {Chen}, {Chen},
  {Chen}, {Chen}, {Cui}, {Cui}, {Deng}, {Dong}, {Du}, {Fu}, {Gao}, {Gao},
  {Gao}, {Gu}, {Guan}, {Gungor}, {Guo}, {Han}, {Huang}, {Huo}, {Jia}, {Jiang},
  {Jiang}, {Jin}, {Kong}, {Li}, {Li}, {Li}, {Li}, {Li}, {Li}, {Li}, {Li}, {Li},
  {Liang}, {Liu}, {Liu}, {Liu}, {Liu}, {Liu}, {Lu}, {Lu}, {Luo}, {Luo}, {Ma},
  {Meng}, {Nang}, {Nie}, {Ou}, {Sai}, {Song}, {Sun}, {Tan}, {Tao}, {Tuo},
  {Wang}, {Wang}, {Wang}, {Wang}, {Wang}, {Wen}, {Wu}, {Wu}, {Wu}, {Xiong},
  {Yang}, {Yang}, {Yang}, {Yang}, {Yin}, {Yin}, {Zhang}, {Zhang}, {Zhang},
  {Zhang}, {Zhang}, {Zhang}, {Zhang}, {Zhang}, {Zhang}, {Zhang}, {Zhang},
  {Zhang}, {Zhang}, {Zhang}, {Zhang}, {Zhang}, {Zhao}, {Zhao}, {Zheng}, {Zhou},
  {Zhu}, \& {Zhu}}]{2021JHEAp}
{Wang}, W., {Tang}, Y.~M., {Ge}, M.~Y., {et~al.} 2021, Journal of High Energy
  Astrophysics, 30, 1.
\newblock \doarXiv{2102.12085}

\bibitem[{{Wang} \& {Frank}(1981)}]{1981A&A....93..255W}
{Wang}, Y.~M., \& {Frank}, J. 1981, \aap, 93, 255

\bibitem[{{Xiao} {et~al.}(2019){Xiao}, {Ji}, {Staubert}, {Ge}, {Zhang},
  {Zhang}, {Santangelo}, {Ducci}, {Liao}, {Guo}, {Li}, {Zhang}, {Qu}, {Lu},
  {Li}, {Song}, {Xu}, {Bu}, {Cai}, {Cao}, {Chang}, {Chen}, {Chen}, {Chen},
  {Chen}, {Chen}, {Chen}, {Cui}, {Cui}, {Deng}, {Dong}, {Du}, {Fu}, {Gao},
  {Gao}, {Gao}, {Gu}, {Guan}, {Gungor}, {Guo}, {Han}, {Huang}, {Huo}, {Jia},
  {Jiang}, {Jiang}, {Jin}, {Kong}, {Li}, {Li}, {Li}, {Li}, {Li}, {Li}, {Li},
  {Li}, {Li}, {Liang}, {Liu}, {Liu}, {Liu}, {Liu}, {Liu}, {Lu}, {Lu}, {Luo},
  {Luo}, {Ma}, {Meng}, {Nang}, {Nie}, {Ou}, {Sai}, {Song}, {Sun}, {Tan}, {Tao},
  {Tuo}, {Wang}, {Wang}, {Wang}, {Wang}, {Wang}, {Wen}, {Wu}, {Wu}, {Wu},
  {Xiong}, {Yang}, {Yang}, {Yang}, {Yang}, {Yin}, {Yin}, {Zhang}, {Zhang},
  {Zhang}, {Zhang}, {Zhang}, {Zhang}, {Zhang}, {Zhang}, {Zhang}, {Zhang},
  {Zhang}, {Zhang}, {Zhang}, {Zhang}, {Zhang}, {Zhang}, {Zhao}, {Zhao},
  {Zheng}, {Zhou}, {Zhu}, \& {Zhu}}]{2019JHEAp..23...29X}
{Xiao}, G.~C., {Ji}, L., {Staubert}, R., {et~al.} 2019, Journal of High Energy
  Astrophysics, 23, 29, \dodoi{10.1016/j.jheap.2019.09.002}

\bibitem[{{Yamamoto} {et~al.}(2014){Yamamoto}, {Mihara}, {Sugizaki},
  {Nakajima}, {Makishima}, \& {Sasano}}]{2014PASJ...66...59Y}
{Yamamoto}, T., {Mihara}, T., {Sugizaki}, M., {et~al.} 2014, \pasj, 66, 59,
  \dodoi{10.1093/pasj/psu028}

\end{thebibliography}
\bibliographystyle{aasjournal}

\end{document}